\documentclass[12pt]{iopart}

\usepackage{iopams}  
\usepackage{xcolor}
\usepackage{ulem}
\begin{document}
\title[Thermodynamic limit of an ideal Bose gas by asymptotic expansions]{The thermodynamic limit of an ideal Bose gas by asymptotic expansions and spectral $\zeta$-functions}
\author{Daniel Alexander Weiss}
\address{Arnold Sommerfeld Center for Theoretical Physics, Theresienstraße 37, 80333 München, Germany}
\ead{weiss.daniel@physik.uni-muenchen.de}
\vspace{10pt}
\begin{indented}
\item[]December 1, 2022
\end{indented}

\begin{abstract}
We analyze the thermodynamic limit – modeled as the open-trap limit of an isotropic harmonic potential – of an ideal, non-relativistic Bose gas with a special emphasis on the phenomenon of Bose-Einstein condensation. This is accomplished by the use of an asymptotic expansion of the grand potential, which is derived by $\zeta$-regularization techniques. Herewith we can show, that the singularity structure of this expansion is directly interwoven with the phase structure of the system: In the non-condensation phase the expansion has a form that resembles usual heat kernel expansions. By this, thermodynamic observables are directly calculable. In contrast, the  expansion exhibits a singularity of infinite order above a critical density and a renormalization of the chemical potential is needed to ensure well-defined thermodynamic observables. Furthermore, the renormalization procedure forces the system to exhibit condensation. In addition, we show that characteristic features of the thermodynamic limit, like the critical density or the internal energy, are entirely encoded in the coefficients of the asymptotic expansion. 
\end{abstract}

\section{Introduction}

It is not often the case that the frontier of contemporary research is present in our daily life. Whenever one cooks water, enjoys snow or uses a magnet, one could ask, how the rich qualitative collective properties of complex systems emerge from the relatively simple properties of its constituents and especially how different macroscopic phases of matter are connected to its microscopic description. But due to the vast complexity of macroscopic systems, the precise answer to those question is unknown in most cases.

A good illustration for the large gap between the understanding of microscopic and the determination of macroscopic phenomena is the phase structure of Bose gases, since, despite of their elementary microscopic description, a rigorous understanding of their phase structure is still lacking in the most realistic scenarios. 

In above lines, Bose gases usually exhibit two phases. One of them is distinguished by a macroscopic occupation of the ground state, what goes under the name of Bose-Einstein condensation (BEC). The corresponding phase transition occurs at low temperatures or high densities and has been experimentally realized in different physical systems (cp. \cite{And95, Cor02}), which is intriguing and fascinating by its own, since it constitutes a macroscopic quantum phenomenon, where the quantum concept of indistinguishability becomes apparent in the macroscopic world. The occurence of BEC was predicted by Bose (\cite{Bose}) and Einstein (\cite{Ein1, Ein2, Ein3}) almost 100 years ago. They analyzed a non-interacting case and argued that the system exhibits a macroscopically occupied ground state below a critical temperature. Nevertheless, a rigorous demonstration of the occurrence of a phase transition in a realistic, interacting scenario was lacking for over 70 years. This changed drastically in 2002, when Lieb and Seiringer proved the occurence of Bose-Einstein condensation in the thermodynamic limit of a dilute Bose gas (\cite{Lie02, Lie05}), which marked a huge progress in the understanding of the phase structure of continuous Bose systems. However, a rigorous understanding in other realistic regimes or for general interactions has still not been achieved. For a good review on this issues see also \cite{Ver10}.

The difficult situation in the interacting case also continuously stimulated research in the much tamer non-interacting case. The hope could be that a new perspective on the non-interacting case also leads to valuable insights concerning the interacting case. Besides the textbook treatment, which uses for example integral approximation techniques (cp. \cite{Bra79}), notable other approaches are the loop gas technique (cp. \cite{Mul00, Bea14}) as well as a recent method which uses insights from algebraic quantum field theory (cp. \cite{Buc22}). Another method for the investigation of the phase structure of ideal bose gases is the method of asymptotic expansions developed in \cite{Kir98, Kir99, Kir10}. In those articles $\zeta$-regularization techniques are used for the investigation of the small-$\beta$ limit of trapped Bose gases. In the present document we will complement those results by an analysis of the thermodynamic or open-trap limit using similar techniques.

Our motivation for the choice of this technique relies on the fact that in other situations the form of such expansions has been proven to be very robust with respect to smooth perturbations of the system. If one considers for example heat kernel expansions on manifolds, the qualitative form of the expansion is insensitive to the geometry of the manifold, and its coefficients are calculable entirely in terms of geometric invariants, which are both non-trivial statements (cp. \cite{Vas03}). Hence, a hope could be that an asymptotic expansion of a characteristic thermodynamic quantity like the grand potential exhibits a similar robustness under perturbations of the system by smooth, repulsive $2$-body interactions and that the occurrence of condensation could be traced back to simple qualitative properties of this expansion. Therefore, a first step is the investigation of the non-interacting case, which is achieved in this document.

For the derivation of the asymptotics of the grand potential under the open-trap limit we utilize $\zeta$-regularization techniques, which are mainly used in finite-temperature relativistic quantum field theory (cp. e.g. \cite{Eli94, Byt03, Eli12}) and are rarely applied to problems in non-relativistic quantum statistical mechanics. In particular, we use the Mellin-Barnes integral representation and the spectral $\zeta$-function of the $1$-particle Hamiltonian in order to extract information on the behavior of the grand partition function under the thermodynamic limit. At this stage we would like to phrase the point, that the utility of spectral $\zeta$-functions in the current situation relies on their capability to translate qualitative properties of the eigenvalue distribution of an operator into precise analytic properties as the resolvents and the pole structure.	

Our paper is organized as follows: After introducing some preliminary notions regarding non-interacting, harmonically trapped Bose gases, in section 3 the small-$\kappa$ asymptotics of the grand potential for the different phase regions corresponding to negative, vanishing and positive renormalized chemical potential is derived. Thereby, we will see that the form of those asymptotic expansions differs drastically in the different phases. In the non-condensation phase the expansion of the grand potential resembles the form of usual Heat Kernel expansions. But in the condensation phase the chemical potential has to be a function of the trap-parameter $\kappa$ and the asymptotic expansion exhibits a singularity of infinite order. In section 4 we will utilize those expansions for the calculation of thermodynamic quantities. We will calculate the average particle density and the average energy density in the non-condensation region in \sref{TQ1} and will derive an expression for the critical density in \sref{CD}. For the analysis of the condensation phase, we draw an analogy to the procedure of renormalization in quantum field theory: By a renormalization of the chemical potential it is possible to cancel the "unphysical" divergencies and render all observables finite. This will be explained in \sref{RenChem}, where also the corresponding renormalization conditions are introduced. In section \ref{RenAs} the conjectured form of the $\kappa$-dependent chemical potential is presented and it is proven, that it satisfies two of the three renormalization conditions. Finally, we will investigate the properties of the condensation phase in \sref{TQ2}. It is shown that the conjectured $\kappa$-dependent chemical potential yields finite results for the considered thermodynamic observables and hence satisfies all renormalization conditions. Furthermore, it is shown that the system exhibits condensation in this regime. Concluding remarks are given in \sref{Conc}.

\section{The non-interacting, harmonically trapped ideal Bose gas}
In this section we will review the necessary prerequisites regarding the quantum statistics of the isotropic harmonic oscillator potential. We consider an ideal Bose gas in $v \geq 1$ dimensions confined to an harmonic oscillator trap. The $1$-particle Hamiltonian is given by
\begin{equation}
T_\kappa := - \Delta + \kappa^2 |\vec{x}|^2
\end{equation}
where $\kappa > 0$ is the considered oscillator constant. The eigenvalues of this operator are given by (cp. \cite{Tes})
\begin{equation}
E^{(v)}_{n_1, ..., n_v}(\kappa) = 2 \kappa \left[\sum_{i=1}^v n_i+ \frac{v}{2} \right]
\end{equation}
for $n_1, ..., n_v \in \mathbb{N}$. The full many-body Hamiltonian is then given by the standard second quantization (cp. \cite{Ara18, Bra79}) 
\begin{equation}
H_\kappa := d\Gamma(T_\kappa) \label{MBH}
\end{equation}
on the bosonic Fock space. The grand canonical potential of the harmonically trapped Bose gas is then given by (cp. \cite{Bra79})
\begin{equation}
\Omega^{(v)}(\kappa; \beta, \mu) = \ln\left[\mathrm{tr}\left(e^{ - \beta (H_\kappa - \mu N) } \right) \right] \label{GCP1}
\end{equation}
where $N$ is the bosonic number operator, $\beta$ is the inverse temperature and $\mu$ is the chemical potential. The expression \eref{GCP1} is well-defined for the parameter ranges $\beta > 0$, $\kappa > 0$ and $- \infty < \mu < E_0^{(v)}(\kappa) $. Here $E_0^{(v)}(\kappa) = \kappa v $ denotes the lowest energy eigenvalue of the $1$-body  Hamiltonian $T_\kappa$. As the starting point of our investigation we will use a sum representation of \eref{GCP1} which is obtained by expanding the logarithm and utilizing a trace formula (cp. Thm. 5.11 of \cite{Ara18}) for second quantized operators on Fock space:
\begin{equation}
\Omega^{(v)}(\kappa; \beta, \mu) = \sum_{n_1, ..., n_v= 0}^{\infty} \sum_{N=1}^\infty \frac{z^N}{N} \exp\left[- \beta N E_{n_1, ..., n_v}^{(v)}(\kappa) \right] \label{ISGC}
\end{equation}
Here we have defined the rapidity $z := e^{\beta \mu}$, that will be used in the following as an equivalent replacement for the chemical potential $\mu$. 
\section{Asymptotic expansions of the grand potential} \label{AEGPS}
We now want to expand the infinite sum \eref{ISGC} in an asymptotic expansion in the trap parameter $\kappa$. This will be realized by utilization of $\zeta$-regularization methods. The starting point is the Mellin-Barnes integral representation (cp. \cite{Kir10})
\begin{equation}
e^{-a} = \frac{1}{2 \pi i} \int_{\sigma - i \infty}^{\sigma + i \infty} a^{-s} \Gamma(s) ds \label{MBI}
\end{equation}
which is valid for $|\mathrm{arg}(a)| < \frac{\pi}{2} - \delta$ with $\delta \in (0, \frac{\pi}{2}]$ and $\sigma > 0$.

 Before we apply the integral formula \eref{MBI} on the sum representation \eref{ISGC}, we want to investigate a simpler situation in order to make the procedure clear (cp. section 6 of \cite{Kir10}).  Consider the more elementary sum:
\begin{equation}
S(\kappa) = \sum_{l=1}^\infty e^{-\kappa l} \label{ElSum1}
\end{equation}
By applying \eref{MBI} we can write \eref{ElSum1} as:
\begin{equation}
S(\kappa) = \sum_{l=1}^\infty \frac{1}{2 \pi i} \int_{\sigma - i \infty}^{\sigma + i \infty} \left(\kappa l\right)^{-s} \Gamma(s) ds \label{Sum1}
\end{equation}
Now recall that the Riemann $\zeta$-function is given for $\mathrm{Re}(s)>1$ by its convergent sum representation (cp. \cite{DLMF}):
\begin{equation}
\zeta_R(s) = \sum_{l=1}^\infty l^{-s}
\end{equation}
Hence, by demanding $\sigma > 1$, we are allowed to interchange the sum and the integral in  \eref{Sum1} and obtain: 
\begin{equation}
S(\kappa) =  \frac{1}{2\pi i} \int_{\sigma - i\infty}^{\sigma + i \infty} \kappa^{-s}\Gamma(s) \zeta_R(s) ds \label{Sum2}
\end{equation}
In analogy to \cite{Kir10}, the strategy for finding the small-$\kappa$ behavior is to shift the integration contour to the left. By the residue theorem, crossing the singularities of the integrand gives then polynomial contributions in $\kappa^{-1}$. In the case of \eref{Sum2}, the rightmost pole of the integrand is given by the pole of $\zeta_R(\cdot)$ at $s=1$ and the other poles can be found at $s = - 2n$ with $n \in \mathbb{N}$. We therefore shift the integral	 contour to $\tilde{\sigma} \in (-1, 0)$ and obtain
\begin{equation}
S(\kappa) = \kappa^{-1}  - \frac{1}{2} + S_{res}(\kappa) \label{Sum3}
\end{equation}
where we used $\mathrm{Res}_{s=1}(\zeta_R(s))= \mathrm{Res}_{s=0}(\Gamma(s)) = 1$ and where the residual term $S_{res}(\kappa)$ is given by:
\begin{equation}
 S_{res}(\kappa)  = \frac{1}{2\pi i} \int_{\gamma(\tilde{\sigma})} \kappa^{-s}\Gamma(s) \zeta_R(s) ds \label{ResTerm} 
\end{equation}
 Here $\gamma(\tilde{\sigma})$ denotes a path in the complex plane which goes from $\sigma -i \infty$ to $\sigma + i \infty$, but intersects the real line at $\tilde{\sigma} \in (-1,0)$. This term gives contributions in $\mathcal{O}(\kappa)$ and is hence of no relevance for us, since we are only interested in the small-$\kappa$ behavior. Thus, we neglect the concrete form of this contribution and obtain the following asymptotic expansion:
\begin{equation}
S(\kappa) = \kappa^{-1} - \frac{1}{2} + \mathcal{O}(\kappa)
\end{equation} 
We now use the same strategy to derive expansions of the form \eref{Sum3} for the grand potential \eref{ISGC}. We will see that we need two different strategies for $\mu \leq 0$ and $\mu > 0$.
\subsection{Negative chemical potential}
We will now derive the small-$\kappa$ asymptotics of the grand potential in the case $\mu < 0$ or equivalently in the case $|z|<1$. Therefore, we apply the Mellin-Barnes integral representation \eref{MBI} on the exponential in \eref{ISGC} and obtain:
\begin{equation}
\Omega^{(v)}(\kappa; \beta, \mu) = \frac{1}{2\pi i} \sum_{n_1, ..., n_v = 0}^\infty \sum_{N=1}^\infty \frac{z^N}{N} \int_{\sigma - i \infty}^{\sigma + i \infty} \beta^{-s} N^{-s} E_{n_1, ..., n_v}^{(v)}(\kappa)^{-s} \Gamma(s) ds \label{ISGC2}
\end{equation}
In contrast to the situation depicted before, the Riemann $\zeta$-function is not sufficient for the analysis of this expression. Instead we need the Barnes $\zeta$-function (see section 2.2 of \cite{Kir10}) whose convergent sum representation is for $v \in \mathbb{N}$, $s>v$, $c > 0$ and $r>0$ given by:
\begin{equation}
\zeta_B^{(v)}(s, c | r) = \sum_{n_1, ..., n_v \in \mathbb{N}_0} \left[c + r \left(n_1 + ... + n_v \right)\right]^{-s}
\end{equation}
If $c = 0$ it is understood, that the sum ranges over $n_1, ..., n_v \neq (0, ..., 0)$. In addition, the polylogarithm is for $|z|< 1$ and any complex order $r \in \mathbb{C}$ given by the following absolute convergent sum (\cite{DLMF}):
\begin{equation}
\mathrm{Li}_r(z) = \sum_{N=1}^\infty \frac{z^N}{N^r} \label{Poly1}
\end{equation}
If we demand $\sigma > v$, we are allowed to interchange the sums and the integral in \eref{ISGC2} and obtain the following expression for the grand potential:
\begin{equation}
\Omega^{(v)}(\kappa; \beta, \mu) = \frac{1}{2\pi i} \int_{\sigma - i \infty}^{\sigma + i \infty} \kappa^{-s} \beta^{-s} \mathrm{Li}_{s+1}(z) \zeta_B^{(v)}(s, v|2) \Gamma(s) ds \label{GPNC}
\end{equation}
The location and residues of the poles for all functions appearing in the integrand are known. In particular, $\mathrm{Li}_{s+1}(z)$ has no poles for $|z|<1$ (cp. \cite{DLMF}) and $\Gamma(s)$ has, as before, simple poles at $- \mathbb{N}_0$. The Barnes $\zeta$-function has poles at $z = 1, ..., v$ (cp. \cite{Kir10}) with residues:
\begin{equation}
\mathrm{Res}_{s = k} \left(\zeta_B^{(v)}(s, c|r) \right) = \frac{(-1)^{v+k}}{(k-1)!(v-k)!} r^{-v} B^{(v)}_{v-k}(c,r)
\end{equation}
Further its value at zero\footnote{We will see in section \ref{RenAs}, that only this value of the Barnes $\zeta$-function is of relevance for us.} is given by (cp. \cite{Kir10}):
\begin{equation}
\zeta_{B}^{(v)}(0,c|r)= \frac{(-1)^v}{v}r^{-v} B_{v}^{(v)}(c,r)
\end{equation}
\noindent Here $B^{(v)}_{m}(c,r)$ denotes generalized Bernoulli polynomials defined as
\begin{equation}
\frac{e^{-xt}}{(1-e^{-rt})^v} = \left(\frac{-1}{r}\right)^v \sum_{n=0}^\infty \frac{(-t)^{n-v}}{n!} B_n^{(v)}(x,r).
\end{equation}

By shifting the contour to the left, we can write now \eref{GPNC} – analogously to \eref{Sum3} – as
\begin{equation}
\Omega^{(v)}(\kappa; \beta, \mu) = \sum_{k= 0}^v \kappa^{-k} a_{-k}^{(v)}(\beta, \mu) + S_{res}(\kappa; \beta, \mu)  \label{SKA-}
\end{equation}
where the coefficients $a_k^{(v)}(\beta, \mu)$ are given by
\begin{equation}
a_{-k}^{(v)}(\beta, \mu) = \cases{\beta^{-k} \mathrm{Li}_{k+1}(z) \Gamma(k) \mathrm{Res}_{s = k} \left(\zeta_B^{(v)}(s, v|2) \right) & $k \in \{1, ..., v\}$ \\\mathrm{Li}_1(z) \zeta_B^{(v)}(0,v|2) & $k = 0$\\} \label{C-1}
\end{equation}
and the residual term $S_{res}(\kappa; \beta, \mu)$ is given by:
\begin{equation}
S_{res}(\kappa) = \frac{1}{2\pi i} \int_{\gamma(\tilde{\sigma})} \kappa^{-s} \beta^{-s} \mathrm{Li}_{s+1}(z) \zeta_B^{(v)}(s, v|2) \Gamma(s) ds \label{Res1}
\end{equation}
Here $\gamma(\tilde{\sigma})$ denotes again a contour that intersects the real axis at $\tilde{\sigma} \in (-1,0)$ and goes from $\sigma -i \infty$ to $\sigma + i \infty$. 
\subsection{Vanishing chemical potential}
The case $\mu = 0$ (or equivalently $z=1$) works analogously to the case $\mu < 0$. The only difference is that the sum over $N$ in \eref{ISGC2} reduces now to a Riemann $\zeta$-function. By this we obtain as an expression for the grand potential in the case $\mu = 0$:
\begin{equation}
\Omega^{(v)}(\kappa; \beta, \mu) = \frac{1}{2\pi i} \int_{\sigma - i \infty}^{\sigma + i \infty } \kappa^{-s} \beta^{-s} \zeta_{R}(s+1) \zeta_B^{(v)}(s,v|2) \Gamma(s) ds
\end{equation}
Here again $\sigma > v$ is required. As before, one shifts the contour to the left and obtains the small-$\kappa$ asymptotics:
\begin{equation}
\Omega^{(v)}(\kappa; \beta, \mu) = \sum_{k= 0}^v \kappa^{-k} a_{-k}^{(v)}(\beta, \mu) + S_{res}(\kappa; \beta, \mu) \label{SKA0}
\end{equation}
The coefficients $a_{-k}^{(v)}(\beta, \mu)$ are then given by
\begin{equation}
a_{-k}^{(v)}(\beta, \mu) = \cases{\beta^{-k} \zeta_R(k+1) \Gamma(k) \mathrm{Res}_{s = k} \left(\zeta_B^{(v)}(s, v|2) \right) & $k \in \{1, ..., v\}$ \\ \mathrm{Res}_{s=0}\left( \zeta_R(s+1) \Gamma(s)\right) \zeta_B^{(v)}(0,v|2) & $k = 0$ \label{C01}}
\end{equation}
and the residual term $S_{res}(\kappa; \beta, \mu)$ is given by:
\begin{equation}
 S_{res}(\kappa; \beta, \mu) = \frac{1}{2\pi i} \int_{\tilde{\sigma} - i \infty}^{\tilde{\sigma} + i \infty } \kappa^{-s} \beta^{-s} \zeta_{R}(s+1) \zeta_B^{(v)}(s,v|2) \Gamma(s) ds \label{Res2}
\end{equation}
$\gamma(\tilde{\sigma})$ denotes again a contour that intersects the real axis at $\tilde{\sigma} \in (-1,0)$ and goes from $\sigma -i \infty$ to $\sigma + i \infty$. 
\subsection{Positive chemical potential}
\label{PRCP}
If the chemical potential in \eref{ISGC} is postive (or equivalently, if $|z|>1$), the previous strategy does not work. The reason for this is twofold. On the one hand, the allowed parameter range for the chemical potential in \eref{GCP1}  is $\mu \in (- \infty, E_0^{(v)}(\kappa))$ and hence one has to choose a $\kappa$-dependent $\mu$, i.e. a map:
\begin{equation}
\mu: \kappa \in (0, \infty) \mapsto \mu(\kappa) \in (0, E_0^{(v)}(\kappa))
\end{equation}
On the other hand, the sum representation of the polylogarithm $\sum_{N=1}^\infty z^NN^{-s-1}$ is in general not convergent for $|z|>1$ and hence one is not allowed to interchange the sums and the integral in \eref{ISGC2}. 

To circumvent this problem, we derive a different sum representation of the grand potential by expanding the exponential that contains the chemical potential. Afterwards we will then apply the same strategy as before on the remaining spectral functions. This gives a small-$\kappa$ asymptotics, where the coefficients are given as power series in the renormalized chemical potential. The motivation for this strategy relies on the observation, that for a positive, $\kappa$-dependent chemical potential the thermodynamic limit corresponds to a weak coupling regime, since $\mu(\kappa) \rightarrow 0$ as $\kappa \rightarrow 0$. When an adequate representation of the $\kappa$-dependent chemical potential as a series in $\kappa$ is given – which will be derived in \sref{RenChem} – this can be reinserted into the expression for the grand potential, yielding again a small-$\kappa$ asymptotics.

Before starting with this program, we have to perform some preliminary steps. First, the occuring spectral functions will be much more convenient, if we perform a redefinition of the chemical potential and the energy eigenvalues by subtracting the zero point energy. We define
\numparts
\begin{eqnarray}
\tilde{\mu}(\kappa) = \mu(\kappa) - E_0^{(v)}(\kappa) \\
\tilde{E}_{n_1, ..., n_v}^{(v)}(\kappa) = E^{(v)}_{n_1, ..., n_v}(\kappa) - E_0^{(v)}(\kappa)
\end{eqnarray}
\endnumparts
and especially we set $\tilde{z}(\kappa) = e^{\beta \tilde{\mu}(\kappa)}$. Please observe, that this implies $\tilde{\mu}(\kappa) < 0$ and $\tilde{z}(\kappa) < 1$. 
Since we are interested in the phenomenon of Bose-Einstein condensation, we further split up the grand potential and separate the ground state contribution
\numparts
\begin{eqnarray}
\Omega^{(v)}(\kappa; \beta, \mu(\kappa)) = \Omega_0^{(v)}(\kappa; \beta, \mu(\kappa)) + \Omega_1^{(v)}(\kappa; \beta, \mu(\kappa))
\end{eqnarray}
where $\Omega^{(v)}_0$ and $\Omega^{(v)}_1$ are given by
\begin{equation}
\Omega_0^{(v)}(\kappa; \beta, \mu(\kappa)) = \sum_{N=1}^\infty N^{-1} \exp\left( \beta N \tilde{\mu}(\kappa) \right)
\end{equation}
and
\begin{equation}
\fl \Omega_1^{(v)}(\kappa; \beta, \mu(\kappa)) =     \sum_{N=1}^\infty   \sum_{(n_1, ..., n_v) \neq (0, ..., 0)} N^{-1} \exp\left(- \beta N ( \tilde{E}_{n_1, ..., n_v}^{(v)}(\kappa) - \tilde{\mu}(\kappa) )\right). \label{EGC1}
\end{equation}
\endnumparts
Since $\tilde{z}(\kappa) < 1$, the ground state contribution can also written as:
\begin{equation}
\Omega_0^{(v)}(\kappa; \beta, \mu(\kappa))  = - \ln\left(1 - \exp\left(\beta \tilde \mu(\kappa) \right)\right)
\end{equation}
For the treatment of $\Omega_1^{(v)}$ we expand
\begin{equation}
\exp\left( \beta N  \tilde{\mu}(\kappa) \right) =\sum_{m=0}^\infty \frac{1}{m!} \beta^m N^m \tilde{\mu}(\kappa)^m
\end{equation}
and insert this into \eref{EGC1}, which gives:
\begin{equation}
  \Omega^{(v)}_1(\kappa; \beta, \mu)  = \sum_{m=0}^\infty \sum_{N=1}^\infty\sum_{(n_1, ..., n_v) \neq (0, ..., 0) } \frac{1}{m!} \beta^m N^{m-1} \tilde{\mu}(\kappa)^m \exp\left(- \beta N \tilde{E}_{n_1, ..., n_v}^{(v)}(\kappa)\right) \label{PEq}
\end{equation}
We now apply as before the Mellin-Barnes integral \eref{MBI}, which gives
\begin{equation}
\Omega^{(v)}_1(\kappa; \beta, \mu) = \sum_{m=0}^\infty \frac{\tilde{\mu}(\kappa)^m }{m!} \int_{\sigma_m -i \infty}^{\sigma_m + i \infty}  \beta^{m-s} \kappa^{-s} \zeta_R(s+1-m) \zeta_B^{(v)}(s, 0|2) \Gamma(s) ds \label{A}
\end{equation}
where  $\sigma_m > \mathrm{max}\{v, m\}$ is required for interchanging the sums over $N$ and $(n_1, ..., n_v)$ in \eref{PEq} with the integrals.

The summands of \eref{A} can then be calculated by the same strategy as before. By shifting the integral contour to $\tilde{\sigma} \in (-1,0)$, one obtains polynomial contributions in $\kappa^{-1}$. As before, the remaining integrals are then collected in $S_{res}(\kappa; \beta, \mu(\kappa))$. By ordering the resulting expression in powers of $\kappa$ we obtain the small-$\kappa$ asymptotics
\begin{equation}
\Omega_1^{(v)}(\kappa; \beta, \mu) = \sum_{k=0}^{\infty} \kappa^{-k} a_{-k}^{(v)}(\beta, \mu(\kappa) ) + S_{res}(\kappa; \beta, \mu(\kappa) )
\end{equation}
where the coefficients $a_k^{(v)}(\beta, \mu(\kappa))$ are given as power-series in $\tilde{\mu}(\kappa)$. Explicitely we obtain for the coefficients:
\numparts
\begin{eqnarray}
 a_0^{(v)}(\beta, \mu(\kappa)) = &- \mathrm{ln}(\beta) + \sum_{m=1}^\infty \frac{\tilde{\mu}(\kappa)^m}{m!} \beta^{m} \zeta_R(1-m)  \zeta^{(v)}_B(0, 0|2) \label{CE1} \\ \nonumber  &+ \mathrm{Res}_{s=0}\left( \zeta_R(s+1) \Gamma(s) \right) \zeta_B^{(v)}(0, 0|2) 
\end{eqnarray}
and
\begin{equation*}
 a_{-k}^{(v)}(\beta, \mu(\kappa) ) = \sum_{m=0, m \neq k}^\infty \frac{\tilde{\mu}(\kappa)^m}{m!} \beta^{m-k} \zeta_R(k+1-m) \mathrm{Res}_{s=k}(\zeta^{(v)}_B(s, 0|2)) \Gamma(k) 
\end{equation*}
\begin{equation} 
 ~~~~~~~~+ \frac{\tilde{\mu}(\kappa)^k}{k!} \Gamma(k) \mathrm{Res}_{s=k}\left( \zeta_R(s+1-k) \zeta_B^{(v)}(s, 0|2)\right)
\end{equation}
for $k =1, ..., v$ and
\begin{equation}
 a_{-k}^{(v)}(\beta, \mu(\kappa) ) = \frac{\tilde{\mu}(\kappa)^k}{k!} \Gamma(k) \zeta_B^{(v)}(k,0|2) \label{CE3}
\end{equation}
\endnumparts
for $k  \geq v+1$. All together we have then the small-$\kappa$ asymptotics 
\numparts
\begin{eqnarray}
\Omega^{(v)}(\kappa; \beta, \mu(\kappa)) = \Omega_0^{(v)}(\kappa; \beta, \mu(\kappa)) + \Omega_1^{(v)}(\kappa; \beta, \mu(\kappa)) \label{IOS0}\\
\Omega_0^{(v)}(\kappa; \beta, \mu(\kappa))  = - \ln\left(1 - \exp\left(\beta \tilde \mu(\kappa) \right)\right) \label{IOS1} \\
\Omega_1^{(v)}(\kappa; \beta, \mu) = \sum_{k=0}^{\infty} \kappa^{-k} a_{-k}^{(v)}(\beta, \mu(\kappa) ) + S_{res}(\kappa; \beta, \mu(\kappa) ) \label{IOS}
\end{eqnarray}
\endnumparts
in the case of positive, $\kappa$-dependent chemical potentials, where the coefficients $a_k^{(v)}(\beta, \mu(\kappa))$ are given by \eref{CE1} -- \eref{CE3} and the residual term is given by:
\begin{equation}
S_{res}(\kappa; \beta, \mu(\kappa) ) = \sum_{m=0}^\infty \frac{\tilde{\mu}(\kappa)^m }{m!} \int_{\gamma(\tilde{\sigma})} \beta^{m-s} \kappa^{-s} \zeta_R(s+1-m) \zeta_B^{(v)}(s, 0|2) \Gamma(s) ds \label{Res3}
\end{equation}
Here $\gamma(\tilde{\sigma})$ denotes as before a contour that intersects the real axis at $\tilde{\sigma} \in (-1,0)$ and goes from $\sigma -i \infty$ to $\sigma + i \infty$. 

At the first glimpse, the apparent singularity of infinite order in the asymptotic expansion \eref{IOS} seems to be unphysical, since it suggests, that also all observables should exhibit a singularity of this type. The resolution of this problem relies on a good choice of the $\kappa$-dependent chemical potential $\mu(\kappa) \in (0, E_0^{(v)}(\kappa) )$. This will be discussed in section \ref{RenChem} and onwards.
\subsection{Summary}
We want to give a short, qualitative summary on the derived asymptotic expansions. In the cases $\mu < 0$ and $\mu = 0$ the asymptotic expansion has the form
\begin{equation}
\Omega^{(v)}(\kappa; \beta, \mu) = \sum_{k=0}^v \kappa^{-k} a_{-k}^{(v)}(\beta, \mu ) + S_{res}(\kappa;\beta, \mu)
\end{equation}
while in the case $\mu > 0$ one has to consider a renormalized chemical potential $\kappa \in (0,\infty) \mapsto \mu(\kappa) \in (0, E^{(v)}_0(\kappa))$ and the asymptotic expansion exhibits a singularity of infinite order:
\begin{equation}
\Omega^{(v)}(\kappa; \beta, \mu(\kappa)) = - \ln\left(1 - \tilde{z}(\kappa) \right) + \sum_{k=0}^{\infty} \kappa^{-k} a_{-k}^{(v)}(\beta, \mu(\kappa) ) +  S_{res}(\kappa;\beta, \mu(\kappa))\label{InfMan}
\end{equation}
The coefficients are given explicitely by \eref{C-1} for the case $\mu < 0$, by \eref{C01} for the case $\mu = 0$ and by \eref{CE1} -- \eref{CE3} for the case of positive, $\kappa$-dependent $\mu$. The forms of the residual terms are also given explicitely by \eref{Res1}, \eref{Res2} and \eref{Res3}.
\section{Behavior under the thermodynamic limit}
We will now utilize the small-$\kappa$ asymptotics of the grand potential as derived in the last section for analyzing the phase structure of the considered Bose gas in the thermodynamic limit $\kappa \rightarrow 0$\footnote{The declaration of the open trap limit $\kappa \rightarrow 0$ as the thermodynamic limit is adequate, since as $\kappa \rightarrow 0$ the average particle number $\langle N \rangle$ diverges.}.
\subsection{Thermodynamic quantities in the non-condensation phase}
\label{TQ1}
We investigate the behavior of some thermodynamic quantities under the limit $\kappa \rightarrow 0$ for fixed, negative chemical potential $\mu < 0$ and arbitrary inverse temperature $\beta > 0$. The average particle number and the average energy are given by their standard expressions:
\numparts
\begin{eqnarray}
N^{(v)}(\kappa; \beta, \mu) &:= \langle N \rangle_{\kappa, \beta, \mu} = \beta^{-1}\frac{\partial}{\partial \mu} \Omega^{(v)}(\kappa; \beta, \mu) \\
E^{(v)}(\kappa; \beta, \mu) &:= \langle H_\kappa \rangle_{\kappa, \beta, \mu} = - \frac{\partial}{\partial \beta} \Omega^{(v)}(\kappa; \beta, \mu)
\end{eqnarray}
\endnumparts
By applying those relations on the small-$\kappa$ asymptotics \eref{SKA-}, we directly obtain small-$\kappa$ asymptotics for those quantities:
\numparts
\begin{eqnarray}
N^{(v)}(\kappa; \beta, \mu) &= \sum_{k=0}^v  n^{(v)}_{-k}(\beta, \mu) \kappa^{-k} + \mathcal{O}(\kappa) \label{APN}\\
E^{(v)}(\kappa; \beta, \mu) &= \sum_{k=0}^v e^{(v)}_{-k}(\beta, \mu) \kappa^{-k} + \mathcal{O}(\kappa) \label{AED}
\end{eqnarray}
\endnumparts
Here the coefficients $n^{(v)}_k(\beta, \mu)$ and $e^{(v)}_k(\beta, \mu)$ are explicitly given by:
\numparts
\begin{eqnarray}
n^{(v)}_{-k}(\beta, \mu) &:= \left( \frac{1}{\beta}\frac{\partial a_{-k}^{(v)}(\beta, \mu)}{\partial \mu} \right) \label{co1} \\
e^{(v)}_{-k}(\beta, \mu) &:= \left(- \frac{\partial a_{-k}^{(v)}(\beta, \mu)}{\partial \beta} \right) \label{co2}
\end{eqnarray}
\endnumparts
By utilizing the identity (\cite{WeiPoly})
\begin{equation}
\frac{d}{dx} \mathrm{Li}_n(x) = \frac{1}{x} \mathrm{Li}_{n-1}(x)
\end{equation}
and applying it on the expressions for the coefficients $a_{-k}^{(v)}(\beta, \mu)$ given in \eref{C-1} we see, that the coefficients $n^{(v)}_{-k}(\beta, \mu)$ and $e^{(v)}_{-k}(\beta, \mu)$ of above expansions are all well-defined and non-zero for $k \in \{0, ..., v\}$. Especially we hence obtain, that the average particle number \eref{APN} and the average energy \eref{AED} exhibit a singularity of order $\kappa^{-v}$. This is not surprising, since in the thermodynamic limit the particle number (and hence also other extensive quantities) should diverge. The meaningful quantities in this regime are hence governed by densities. For this we consider the inverse trap parameter $\kappa^{-1}$ as the characteristic length scale of the problem, which makes it plausible to think of $\kappa^{-v}$ as the characteristic volume of the harmonic trap (for this viewpoint, see also \cite{Mul00, Bea14}). This motivates us to define the particle- and the energy density as\footnote{A more physical approach would have been to consider the quotient $\tilde{\rho}^{(v)}_E(\kappa; \beta, \mu) = \langle N \rangle^{-1} \langle E \rangle$. But since $\langle N \rangle$ diverges as $\kappa^{-v}$, this would have been qualitatively the same and would just correspond, in the light of section \ref{RenChem}, to a change of the renormalization point.}:
\numparts
\begin{eqnarray}
\rho^{(v)}(\kappa; \beta, \mu) := \kappa^v N^{(v)}(\kappa; \beta, \mu)\\
\rho^{(v)}_E(\kappa; \beta, \mu) := \kappa^v E^{(v)}(\kappa; \beta, \mu)
\end{eqnarray}
\endnumparts
We see then, that under the thermodynamic limit $\kappa \rightarrow 0$ the expressions for the average density and the average energy are directly given by the coefficients $n^{(v)}_{-v}(\beta, \mu)$ and $e^{(v)}_{-v}(\beta, \mu)$ in \eref{co1} and \eref{co2}. If one inserts the expressions for the coefficients $a_{-v}^{(v)}$ one then obtains:
\numparts
\begin{eqnarray}
 \rho^{(v)}(\beta, \mu) := \lim_{\kappa \rightarrow \infty} \rho^{(v)}(\kappa; \beta, \mu) = 2^{-v} \beta^{-v}   \mathrm{Li}_v(z) \label{R1}\\
 \rho^{(v)}_E(\beta, \mu) :=  \lim_{\kappa \rightarrow \infty} \rho^{(v)}_E(\kappa; \beta, \mu)  = v 2^{-v} \beta^{-v-1} \mathrm{Li}_{v+1}(z)  
\end{eqnarray}
\endnumparts
Finally we want to show, that no condensation occurs for $\mu \leq 0$ . Therefore consider the average ground-state occupation density given by (cp. \cite{Kir99}):
\begin{equation}
\rho_0^{(v)}(\kappa; \beta, \mu) = \frac{\kappa^v}{\beta} \frac{\partial}{\partial \mu} \Omega_0^{(v)}(\kappa; \beta, \mu) = \kappa^v \left(1- \exp[-\beta( E_0^{(v)}(\kappa)- \mu)] \right)
\end{equation}
We then see, that $\rho_0^{(v)}(\kappa; \beta, \mu) \rightarrow 0$ as $\kappa \rightarrow 0$ for $\mu \leq 0$. 
\subsection{The critical density}
\label{CD}
In the last section we have analyzed the thermodynamic limit of an ideal Bose gas for $\mu \leq 0$. By observing, that the function
\begin{equation}
\mu \in (- \infty, 0) \mapsto \rho^{(v)}(\beta, \mu)
\end{equation}
is strictly monotonically increasing, we can easily calculate the maximal density $\rho_{c}^{(v)}(\beta)$ that can be attained in this phase:
\begin{equation}
\rho_{c}^{(v)}(\beta) := \sup_{\mu < 0} \rho^{(v)}(\beta, \mu)= \lim_{\mu \rightarrow 0} \rho^{(v)}(\beta, \mu)
\end{equation}
This maximal density will be called the critical density. By recalling $\mathrm{Li}_v(1) = \zeta_R(v)$ (cp. \cite{WeiPoly}) we then obtain:
\begin{equation}
\rho_{c}^{(v)}(\beta) = 2^{-v} \beta^{-v} \zeta_R(v)
\end{equation}
which is finite for $v \geq 2$, infinite for $v = 1$ and is equivalent to the results of \cite{Bea14} if one takes their different conventions into account. Hence one has to choose an adequate positive, $\kappa$-dependent chemical potential to obtain higher densities.  This causes the system to exhibit condensation as we will show \sref{TQ2}.
\subsection{Renormalization of the chemical potential}
\label{RenChem}
Let from now on $v \geq 2$. We have seen, that the critical density $\rho_{c}^{(v)}(\beta)$ is finite in this case and hence one has to use a $\kappa$-dependent, positive chemical potential to obtain higher densities $\bar{\rho} > \rho_{c}^{(v)}(\beta)$. If one recalls the form of the small-$\kappa$ asymptotics \eref{InfMan} it is a priori not clear, that there exist such $\kappa$-dependent chemical potentials, which yield meaningful results for thermodynamic observables in the limit $\kappa \rightarrow 0$. Nevertheless, as discussed before, this situation corresponds to a weak coupling regime, where the $\kappa$-dependent chemical potential satisfies $\mu(\kappa) \in (0, E^{(v)}_0(\kappa) )$ with the zero-point energy behaving as $E^{(v)}_0(\kappa) \rightarrow 0$ as $\kappa \rightarrow 0$. This suggests, that taking the thermodynamic limit corresponds to the consideration of arbitrary small neighborhoods around $\mu = 0$, and hence the asymptotic expansions obtained for $\kappa \rightarrow 0$ should not differ so drastically in the two cases $\mu \leq 0$ and $\mu > 0$. Our strategy will now be to analyze, if there exist such $\kappa$-dependent chemical potentials, which yield a finite density $\bar{\rho}$ in the limit $\kappa \rightarrow 0$ and which regularize the small-$\kappa$ asymptotics, such that it exhibits only a singularity of finite order. More precisely stated, the question is hence, if there exists for any $\bar{\rho} > \rho_{c}^{(v)}(\beta)$ a $\kappa$-dependent chemical potential
\begin{equation}
\mu_{\bar{\rho}}: \kappa \in (- \infty, 0) \mapsto \mu_{\bar{\rho}}(\kappa) \in (0, E_0^{(v)}(\kappa))
\end{equation}
which satisfies the following conditions, which will be called renormalization conditions in the sequel:
\begin{enumerate}
\item $\lim_{\kappa \rightarrow 0} \mu_{\bar{\rho}}(\kappa) = 0$.
\item $\lim_{\kappa \rightarrow 0} \rho^{(v)}\left(\beta, \mu_{\bar{\rho}}(\kappa) \right) = \bar{\rho}$.
\item For all $k > v$ it holds, that the coefficient $a_{-k}^{(v)}(\beta, \mu_{\bar \rho}(\kappa))$ in \eref{InfMan} lies in $\mathcal{O}(\kappa^{k+1})$. 
\end{enumerate}
Before we show, that there exist such $\mu_{\bar{\rho}}(\kappa)$, we would like to draw an analogy to the renormalization procedure in quantum field theory, to justify, why we call this procedure a renormalization of the chemical potential. Therefore recall, how observables are calculated in quantum field theory (cp. e.g. \cite{Itz06}): A theory is specified by a Lagrangian, which contains several microscopic ("bare") parameters, like the mass or the coupling. If one tries naivly to calculate observables by the evaluation of Feynman diagrams, one obtains divergent integrals. To cure this problems, one follows a two-step strategy. First, an UV-regulator, say a cut-off frequency $\Lambda$, is introduced to render the observables finite. In a second step, which is called the renormalization, the microscopic parameters are chosen to be regulator-dependent in a way specified by certain renormalization conditions, which ensure, that all occuring divergencies in the observables are canceled. After this procedure, the theory is reparametrized: The parameters of the theory are not given by the microscopic parameters anymore, but by renormalized, physical parameters determined by the remaining degrees of freedom of the renormalization prescription. In so called on-shell schemes, for example, the renormalized parameters are explicitely defined as the outcomes of certain diagrams, that can be interpreted as measurements.

Observe now, that our situation is quite similar. Therefore one has to view the trap-parameter $\kappa$ as an IR-regulator and the chemical potential $\mu$ as a microscopic ("bare") parameter of the system. If one tries then to calculate observables in the condensation phase ($\mu > 0$) in a naive way, one obtains, that all observables diverge as the IR-regulator is removed (i.e. as $\kappa \rightarrow 0$). The reason for this is, that the grand potential exhibits a singularity of infinite order, as it is shown by the asymptotic expansion \eref{IOS}. Our strategy is then, to choose a regulator-dependent (i.e. $\kappa$-dependent) chemical potential $\mu(\kappa)$, which cancels all occuring divergencies in the observables. As before, the precise form of this regulator-dependence is determined by certain renormalization conditions, namely in our case by above conditions (i) - (iii). After this renormalization procedure, the observables do not depend on a microscopic parameter $\mu$ anymore (especially we see, that a fixed, positive $\mu$ is not a meaningful parameter in the condensation phase), but on the macroscopic parameter $\bar{\rho}$ defined by the renormalization prescription. In our case, this parameter is defined as the outcome of a density measurement and could be called, in the light of above analogy, the renormalized chemical potential. Hence we see, that a choice of a $\kappa$-dependent $\mu$, as determined by above conditions (i) - (iii), resembles the procedure of renormalization in high energy physics, and and therefore we call this procedure a renormalization of the chemical potential and the above conditions renormalization conditions.

In the next two sections we will perform this procedure. Therefore we will show in section \ref{RenAs}, that $\kappa$-dependent chemical potentials satisfying the renormalization conditions (i) and (iii) exist and hence cure the singularity of infinite order in the asymptotic expansion of the grand potential. In section \ref{TQ2} we will then show, that those $\kappa$-dependent $\mu$ satisfy the second renormalization condition and calculate additional thermodynamic observables.

\subsection{Renormalized asymptotic expansions in the condensation phase}
\label{RenAs}
The first renormalization condition is trivially satisfied, since $\mu_{\bar{\rho}}(\kappa) \in (0, E_0^{(v)}(\kappa))$, while the last renormalization condition ensures, that the small-$\kappa$ asymptotics exhibits only finitely many singularities. We show in this subsection, that there exists a renormalized chemical potential $\mu_{\bar{\rho}}(\kappa)$, which satisfies the first and the third renormalization condition. The second renormalization condition will then be shown in the next section.

We guess the form of the $\kappa$-dependent chemical potential as:
\begin{equation}
\mu_{\bar{\rho}}(\kappa) = E_0^{(v)}(\kappa) - \left[ \bar{\rho} - \rho_{c}^{(v)}(\beta) \right]^{-1} \kappa^{v} + \mathcal{O}(\kappa^{v+1}) \label{RCP}
\end{equation}
Since $E_0^{(v)}(\kappa) \rightarrow 0$ as $\kappa \rightarrow 0$, the $\kappa$-dependent chemical potential \eref{RCP} trivially satisfies the first renormalization condition. We have further, by recalling \eref{CE3},  that $a_{-k}(\beta, \mu(\kappa))$ is for $k>v$ given by
\begin{equation}
a_{-k}(\beta, \mu_{\bar{\rho}}(\kappa)) = k^{-1} \left( \mu_{\bar{\rho}}(\kappa) - E_0^{(v)}(\kappa) \right)^k \zeta_B^{(v)}(k, 0|2) 
\end{equation}
and hence lies in $\mathcal{O}(\kappa^{vk})$. Consequently, the first and the third renormalization condition are satisfied. By this we obtain that the form of the asymptotic expansion of the grand potential attains the following form, if a renormalized chemical potential \eref{RCP} is inserted and if one assumes, that the residual term lies in $\mathcal{O}(\kappa)$:
\numparts
\begin{eqnarray}
\Omega^{(v)}(\kappa; \beta, \mu_{\bar \rho}(\kappa)) = \Omega_0^{(v)}(\kappa; \beta, \mu_{\bar \rho}(\kappa)) + \Omega_1^{(v)}(\kappa; \beta, \mu_{\bar \rho}(\kappa)) \label{SKR1}\\
\Omega_0^{(v)}(\kappa; \beta, \mu_{\rho}(\kappa)) = - \ln(1- \tilde{z}_{\bar{\rho}}(\kappa) ) \\
\Omega_1^{(v)}(\kappa; \beta, \mu_{\bar \rho}(\kappa) ) = \sum_{k=0}^v \kappa^{-k} a^{(v)}_{-k}(\beta) + \mathcal{O}(\kappa) \label{SKR3}
\end{eqnarray}
\endnumparts
Here the coefficients $a_{-k}^{(v)}(\beta)$ are given by
\numparts
\begin{equation}
a_{-k}^{(v)}(\beta) = \beta^{-k} \zeta_R(k+1) \Gamma(k) \mathrm{Res}_{s=k}\left( \zeta_B(s,0|2)\right) \label{SKRC1}
\end{equation}
for $k = 1, ..., v$ and
\begin{equation}
a_0^{(v)}(\beta) = - \ln(\beta) \label{SKRC2}
\end{equation}
\endnumparts
for $k=0$, where we dropped the contributions in $\mathcal{O}(\kappa)$. We hence see, that the small-$\kappa$ asymptotics of $\Omega_1^{(v)}$ with renormalized chemical potential almost completely resembles the form of the small-$\kappa$ asymptotics in the case $\mu = 0$. 
The second renormalization condition will be investigated in the next subsection.
\subsection{Thermodynamic quantities in the condensation phase}
\label{TQ2}
We now perform an analogous analysis as in \sref{TQ1}. Thereby we show, that the renormalized chemical potential \eref{RCP} yields finite expressions for the considered thermodynamic quantities and hence especially satisfies the second renormalization condition from \sref{RenChem}. As in section \sref{TQ1} the average particle density and the average energy density are given by:
\numparts
\begin{eqnarray}
\rho^{(v)}(\kappa; \beta, \mu_{\bar{\rho}}(\kappa)) = \kappa^v \beta^{-1} \frac{\partial}{\partial \mu} \Omega^{(v)}(\kappa; \beta, \mu_{\bar{\rho}}(\kappa) ) \\
\rho^{(v)}_E(\kappa; \beta, \mu_{\bar{\rho}}(\kappa)) = - \kappa^v  \frac{\partial}{\partial \beta} \Omega^{(v)}(\kappa; \beta, \mu_{\bar{\rho}}(\kappa) ) 
\end{eqnarray}
\endnumparts
If one inserts the small-$\kappa$ asymptotics \eref{SKR1} -- \eref{SKR3} together with the expressions for the coefficients \eref{SKRC1} and \eref{SKRC2} in those expressions, one obtains the following expression for the densities in the thermodynamic limit:
\begin{eqnarray}
\rho^{(v)}(\beta, \bar{\rho}) := \lim_{\kappa \rightarrow 0} \rho^{(v)}(\kappa; \beta, \mu_{\bar{\rho}}(\kappa)) = \bar{\rho}\\
\rho^{(v)}_E(\beta, \bar{\rho}) := \lim_{\kappa \rightarrow 0} \rho^{(v)}_E(\kappa; \beta, \mu_{\bar{\rho}}(\kappa)) = \rho_E^{(v)}(\beta, 0)
\end{eqnarray}
Finally we want to analyze, if the system exhibits condensation. As in \sref{TQ1} the expression for the average ground-state occupation density is given by
\begin{equation}
\rho_0^{(v)}(\kappa; \beta, \mu_{\bar{\rho}}(\kappa)) = \frac{\kappa^v}{\beta} \frac{\partial}{\partial \mu} \Omega_0^{(v)}(\kappa; \beta, \mu_{\bar{\rho}}(\kappa))
\end{equation}
and one obtains by a direct calculation:
\begin{equation}
\lim_{\kappa \rightarrow 0}\rho_0^{(v)}(\kappa; \beta, \mu) = \bar{\rho} - \rho_{c}^{(v)}(\beta)
\end{equation}
Hence the system exhibits condensation. 
\section{Conclusion}
\label{Conc}
In this article we have presented a detailed analysis of the thermodynamic or open-trap limit of harmonically trapped Bose gases by the method of asymptotic expansions. In particular, our results for the renormalized chemical potential and the critical density are equivalent to the results of \cite{Bea14} if one takes their different conventions into account. A natural question is now, if the current method is also of any use in the case of elementary interactions. For this one has to analyze, how a perturbation of the energy-spectrum alters the analytic properties of the occuring spectral functions. A first step in this direction would be to consider elementary interactions – as repulsive contact-interactions or rapidly decaying $2$-body potentials – in the present framework up to first order in Rayleigh-Schrödinger perturbation theory. One could then analyze the weak-coupling regime analogously to the analysis of \sref{PRCP} by trying to separate the spectral $\zeta$-functions of the unperturbed problem from the contributions caused by the perturbation. Maybe it could be possible to obtain true asymptotic series for the grand partition function, where the coefficients are then not given as series in the renormalized chemical potential alone, but also as series in the coupling $\lambda$ of the interaction Hamiltonian. But this analysis lies beyond the scope of the current article.
\appendix

\section*{Acknowledgements}
I would like to thank Wojciech Dybalski for fruitful discussions and the Norbert Janssen foundation for financial support. 

\section*{Data availability statement}
The data that supports the findings of this study are available within the article.
\newpage 
\section*{References}
\bibliography{bibliography-b}

\end{document}